\documentclass[aps,prl,floats,superscriptaddress,showpacs,twocolumn]{revtex4-1}

\usepackage{amsmath, amsthm, amssymb}
\usepackage{enumitem}
\usepackage{graphics,graphicx}% Include figure files
\usepackage{epsfig}
\usepackage{times}
\usepackage{dcolumn}% Align table columns on decimal point
\usepackage{bm}% bold math

\usepackage{ulem, color}

\begin{document}

\title{Experimental implementation of maximally synchronizable networks}

\author{R. Sevilla-Escoboza}
\affiliation{Centro Universitario de los Lagos, Universidad de Guadalajara, Enrique D\'{i}az de Leon, Paseos de la Monta\~na, Lagos de Moreno, Jalisco 47460, Mexico}
\author{J. M. Buld\'u}
\affiliation{Laboratory of Biological Networks, Center for Biomedical Technology, Technical University of Madrid, Pozuelo de Alarc\'{o}n, 28223 Madrid, Spain}
\affiliation{Complex Systems Group \& GISC, Universidad Rey Juan Carlos, 28933 M\'ostoles, Madrid, Spain}
\author{S. Boccaletti}
\affiliation{CNR-Istituto dei Sistemi Complessi, Via Madonna del Piano, 10, 50019 Sesto Fiorentino, Italy}
\affiliation{The Italian Embassy in Israel, 25 Hamered Street, 68125 Tel Aviv, Israel}
\author{D. Papo}
\affiliation{Laboratory of Biological Networks, Center for Biomedical Technology, Technical University of Madrid, Pozuelo de Alarc\'{o}n, 28223 Madrid, Spain}
\author{D.-U. Hwang}
\affiliation{National Institute for Mathematical Sciences, Daejeon 305-811, South Korea}
\author{G. Huerta-Cuellar}
\affiliation{Centro Universitario de los Lagos, Universidad de Guadalajara, Enrique D\'{i}az de Leon, Paseos de la Monta\~na, Lagos de Moreno, Jalisco 47460, Mexico}
\author{R. Guti\'errez}
\affiliation{Department of Chemical Physics, The Weizmann Institute of Science, Rehovot 76100, Israel}
\affiliation{School of Physics and Astronomy, University of Nottingham, Nottingham NG7 2RD, United Kingdom}

%%%%%%%%%%%%%%%%%%%%%%%%%%%%%%%%%%%%%%%%%%%%%%%%%%%%%%%%%%%%%%%%%%%

%%%%%%%%%%%%%%%%%%%%%%%%%%%%%%%%%%%%%%%%%%%%%%%%%%%%%%%%%%%%%%%%%%%
\begin{abstract}
%%%%%%%%%%%%%%%%%%%%%%%%%%%%%%%%%%%%%%%%%%%%%%%%%%%%%%%%%%%%%%%%%%%
Maximally synchronizable networks (MSNs) are acyclic directed networks that maximize synchronizability. In this paper, we investigate the feasibility of transforming networks of coupled oscillators into their corresponding MSNs. By tuning the weights of any given network so as to reach the lowest possible eigenratio $\lambda_N/\lambda_2$, the synchronized state is guaranteed to be maintained across the longest possible range of coupling strengths. We check the robustness of the resulting MSNs with an experimental implementation of a network of nonlinear electronic oscillators and study the propagation of the synchronization errors through the network. Importantly, a method to study the effects of topological uncertainties on the synchronizability is proposed and explored both theoretically and experimentally.

\end{abstract}

\maketitle

\section{Introduction}

Synchronization is a paradigmatic example of collective behavior in physical and biological sciences. The scientific study of synchronization started with the pioneering observations on the dynamics of pendulum clocks hanging from a common beam by Christiaan Huygens in the 17th century \cite{pikovsky2001}, and continue these days with studies showing different forms of synchronization in networks of both regular and chaotic self-oscillating dynamics. Throughout the years, evidence of synchronized behavior has been found in mechanical, electronic and neuronal systems, as well as in chemical reactions and biological rhythms, to name but a few examples. A flurry of interest in synchronization started in the mid 1990s, with some earlier breakthroughs such as the Kuramoto model \cite{kuramoto1984} paving the ground for it. The results of those fruitful years of research are reviewed in \cite{pikovsky2001,boccaletti2002}.

Since the beginning of the previous decade, the focus has been moving from the study of the synchronized dynamics of just two or a few dynamical system, towards the study or large ensembles of oscillators with complex coupling arrangements. In this regard, the development of network theory \cite{boccaletti2006,newman2010} allowed for a fruitful interaction between a dynamical point of view and a more topological perspective that has led to the present-day field of the synchronization of complex networks \cite{boccaletti2006,arenas2008}.

Among these new developments, the {\it master stability function} (MSF) approach \cite{pecora1998,barahona2002} is a cornerstone of network synchronization research, and it provides the language in which the ideas we develop are written. The MSF describes the stability of the synchronized state of a set of $N$ coupled dynamical units as a function of the coupling parameter $\sigma$ and the topology of the network under study. Given a network of $N$ dynamical systems, the evolution (in isolation) of the state of each node is described by a set of $n$-dimensional differential equations ${\bf \dot x}_i={\bf F}({\bf x}_i)$, where \boldsymbol{${x}_i \in {R^n}$} is the dynamic state vector of node $i$. If nodes have a certain interaction between them, the evolution of the coupled systems is given by:
\begin{eqnarray}
{\bf{\dot x}_i}={\bf F}({\bf x}_i) - \sigma \sum^N_{j=1} \mathcal{L}_{ij} {\bf H(x}_j), \;\;\;\;\;\;\;\;\;  i=1,\ldots,N
\label{eq:msf}
\end{eqnarray}
where $\sigma$ is the coupling strength, \boldsymbol{${\bf H}({\bf x}): {R}^n \rightarrow {R}^n$} is a vectorial coupling function and
$\mathcal{L}_{ij}$ are the elements of the network Laplacian matrix. We here consider weighted networks: the connectivity is given by a weighted adjacency matrix whose elements $W_{ij}$ are real numbers giving the link weights between nodes $i$ and $j$ if they are connected and zero otherwise. Equivalently, the coupling can be represented (as in Eq. \ref{eq:msf}) by a weighted Laplacian matrix defined as $\mathcal{L}_{ij} = \delta_{ij} \sum_k W_{ik} - W_{ij}$. Note that Eq. (\ref{eq:msf}) is equivalent to a diffusive coupling between the nodes of the network, the weight of the coupling contained in $W_{ij}$. Due to the zero-row sum property of the Laplacian matrix $(\sum_j \mathcal{L}_{ij}=0)$ the synchronization manifold ${\bf x}_1={\bf x}_2=...={\bf x}_N\equiv{\bf x}_s$, with ${\bf \dot x}_s=F({\bf x}_s)$, is an invariant set of the dynamics. Under these conditions, the MSF approach provides a framework for the study of the stability of synchronization in which the topology and the dynamics are in some sense uncoupled. The stability of the synchronized dynamics or {\it synchronizability} is established by computing the maximum Lyapunov exponent of a suitably modified kernel that depends on a parameter $\nu$, which is proportional to the coupling strength in the network $\nu = \sigma \lambda$ \cite{pecora1998}. This Lyapunov exponent can be seen as the parameter giving the exponential divergence/convergence of perturbations orthogonal to the synchronization manifold, and when parameterized in terms of $\nu$ gives the MSF curve, which we denote as $\Lambda(\nu)$. The proportionality constant $\lambda$ represents one of the nonzero eigenvalues of the graph Laplacian matrix. The graph Laplacian matrices of the networks considered in this work have a real and non-negative spectrum, with eigenvalues $0 = \lambda_1 < \lambda_2 \leq \lambda_3 \leq \cdots \leq \lambda_N$, where $N$ is the number of nodes (dynamical units) of the network. Moreover, as the networks are connected, only one eigenvalue is zero, and thus the \textcolor{black}{$\nu$} corresponding to the different oscillation eigenmodes will be positive. For those values of $\nu$ for which the MSF is negative, perturbations transversal to the synchronization manifold damp out exponentially fast. For an uncoupled network, $\nu = 0$, the MSF corresponds to that of the autonomous dynamics at each node, and is therefore either zero (for regular dynamics) or positive (for chaotic dynamics). As $\nu$ is varied, the MSF may become negative in some  regions, and it is of specially interest to study the boundaries of those regions. When $\Lambda(\nu)$ is negative for all \textcolor{black}{$\sigma \lambda_i$ with $i = 2, 3, \ldots, N$}, the synchronization manifold is stable and the network is said to be synchronizable \cite{pecora1998}.

For a given dynamical system and coupling function, three categories can be defined according to the classification proposed in Ref. \cite{boccaletti2006}: Class I systems have a non-negative MSF (and are therefore not synchronizable), class II systems are those that have an unbounded synchronizability region ($\Lambda(\nu)<0$ for $\nu > \nu_c$, where $\nu_c$ represents the only zero of the MSF) and class III systems are those that have a bounded synchronizability region ($\Lambda(\nu)<0$ for $\nu \in (\nu_1, \nu_2)$, where $\nu_1$ and $\nu_2$ are the two zeros of the MSF). In practice, only a finite range of $\nu$ that is expected to cover all the cases of interest is considered, and the classification is applied in this restricted sense. If one considers larger $\nu$ ranges, MSFs with more than two zero crossings are indeed possible, and some complicated MSF curves have been reported for example in \cite{huang2009}.

Of these three main classes, class I and class II are, in some sense, trivial. Class I is not synchronizable, while class II systems are always synchronizable for large enough $\sigma$, specifically for $\sigma > \nu_c/\lambda_2$. However, the conditions for class III systems to be synchronizable, $\lambda_2 \sigma > \nu_1$ and $\lambda_N \sigma < \nu_2$, imply that a topology such that $\lambda_N/\lambda_2 > \nu_2/\nu_1$ is not synchronizable for any value of $\sigma$. In fact, whatever the system, if it is class III, a topology with a smaller eigenratio $R \equiv \lambda_N/\lambda_2$ is easier to synchronize as the different $\nu_2 \leq \cdots \leq \nu_N$ that have to be accommodated within the synchronization region are bunched together more closely. Following this argument, the optimal case is that for which $\lambda_2 = \lambda_3 = \cdots = \lambda_N$, in which all the \textcolor{black}{relevant values of $\nu$ for the given topology, namely $\sigma \lambda_i$} for $i=2,3,\ldots,N$, become equal, and the eigenratio $R$ reaches its minimum $R = 1$, leading to an optimal synchronizability (i.e., the one being stable for a larger range of $\sigma$). Networks with this particular structure are known as  {\it maximally synchronizable networks} (MSN) \cite{nishikawa2006}.

In this paper, we investigate the feasibility of transforming a given network into its MSN and test its robustness to noise and parameter mismatch in real systems. While some publications have shown the possibility of enhancing the synchronizability of networks by rearranging the links \cite{motter2005,chavez2005}, in many cases the creation or deletion of links is not available due to experimental restriction. Here, we consider the architecture underlying the network topology to be fixed (nodes are connected or disconnected once and for all), and try to achieve the graph that optimizes the stability of the synchronized state by tuning the link weights. We design an experimental implementation, by means of electronic circuits, of the MSN and investigate its stability as a function of the coupling strength. Interestingly, we observe how the synchronization error is propagated through the network when the system is close to the synchronization boundaries. Next, we analyze the effects of the deviations from the optimal topology on the synchronization of the whole system, and show the interplay between the {\it topological noise} with the coupling strength of the whole network.

\section{Maximally synchronizable networks: experimental implementation}

\textcolor{black}{In this section we discuss the experimental implementation of MSNs and the corresponding results. It is divided into three subsections: in the first one, we describe the algorithm that produces MSNs out of arbitrary topologies; then we describe the dynamical system of use, and the experimental setup based on electronic circuits; finally, we discuss the experimental results.}

\subsection{\textcolor{black}{Maximally synchronizable network algorithm}}

Any undirected (connected) network can be converted into a MSN by the following procedure \cite{nishikawa2006}:

\begin{enumerate}
\item Select any node of the network as the initial node (from now on, node $1$).
\item Number the $k_1$ neighbors of node $1$ sequentially (i.e., give the numers $2, 3, \ldots, k_1+1$ to them).
\item Repeat the process with the second neighbors (i.e. the neighbors of $2, 3, \ldots, k_1+1$), third neighbors (i.e. the neighbors of the neighbors of $2, 3, \ldots, k_1+1$). and so on, until we have numbered to all the nodes of the network.
\item Transform the links into unidirectional links pointing from the node that has been assigned the lower number to the node with the greater number.
\item Give a weight of $1/k^{in}_i$ to all incoming links of nodes $i = 2,\ldots,N$, $k^{in}_i$ being the in-degree of node $i$ (i.e. the number of neighbors whose links point to it), so that the total incoming weight of all nodes will be one, with the exception of node 1 (which is zero).
\end{enumerate}

\begin{figure}
\begin{center}
	
\includegraphics[scale=0.10]{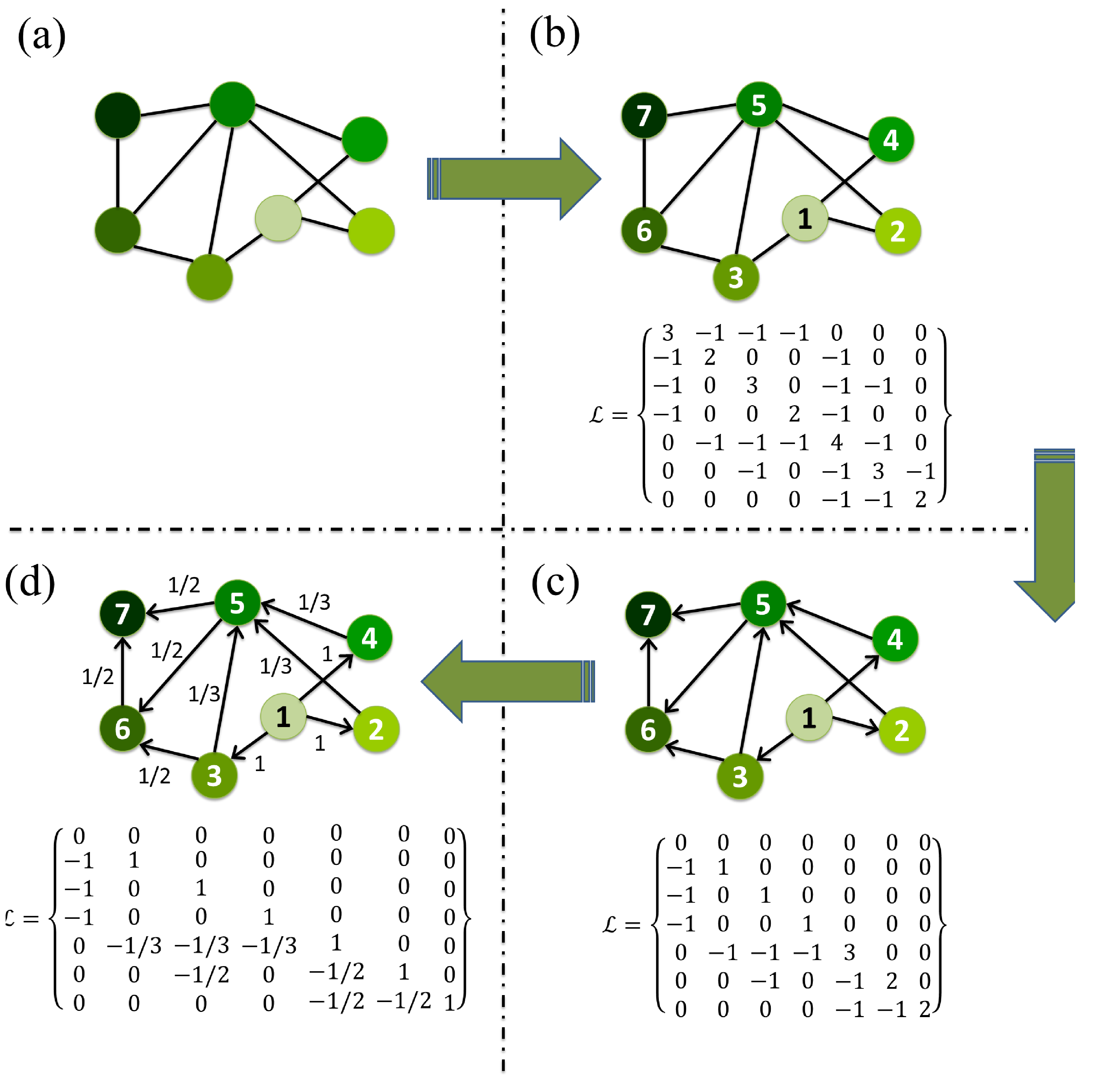}

\end{center}
\caption{ {\sf \bf Construction of Maximally Synchronizable Network (MSN).}
(a) Starting from a weighted undirected network, (b) the first step is the numbering process, which consists in
selecting an initial node and sequentially numbering its first neighbours. The process is repeated for successive layers of neighbors until the whole network has been numbered. (c) Next, directions are given to the links with arrows pointing
from the node with the lower number to the node with the higher number. (d) Finally, the weights of the links pointing to node $i$ are set to $1/k^{in}_i$.
}\label{fig1}
\end{figure}

By following these steps, we obtain a directed weighted network with a weighted adjacency matrix such that $W_{ij} = 1/{k^{in}_i}$ if there is a link $(j \to i)$ or zero otherwise. \textcolor{black}{The procedure is equally applicable whether the original network is weighted or not, as only the basic structure as given by the unweighted (binary) adjacency matrix is used.} The corresponding elements of the resulting Laplacian matrix are
\begin{equation}
\mathcal{L}_{ij} = \left\{
  \begin{array}{lcc}
    $1$ 			& \mbox{if $i=j>1$}\\
    -1/{k^{in}_i} & \mbox{if $j<i$ and $i$ and $j$ are connected}\\
	$0$ 			& \mbox{otherwise}\\
  \end{array} \right.
\end{equation}
The full Laplacian matrix therefore is a lower triangular matrix
\begin{equation}
\mathcal{L} = \left(
 \begin{array}{l l l l l}
  0 & 0 &  0 &\cdots & 0 \\
  -1 & 1 & 0  & \cdots & 0 \\
  -1/k^{in}_3 (0) & -1/k^{in}_3 (0) & 1 & \cdots & 0 \\
  \vdots  & \vdots & \vdots & \ddots & \vdots  \\
  -1/k^{in}_N (0) & -1/k^{in}_N (0) & -1/k^{in}_N (0) & \cdots & 1
 \end{array} \right )
\end{equation}
where the element $\mathcal{L}_{ij}$ for $j < i$ being denoted as $-1/k^{in}_i (0)$ means that it can be either $-1/k^{in}_i$ if there is a link from $j$ to $i$ or zero if there is not. As the network is connected, node $1$ must have at least one neighbor, and therefore there must be a link from $1$ to $2$, and that is why we do not apply this notation to $\mathcal{L}_{21}$, which is known to be $-1$. The procedure and the resulting Laplacian matrix are illustrated in Fig. \ref{fig1}. \textcolor{black}{As the spectrum of a triangular matrix is given by the elements along the main diagonal, it is obvious that the matrix corresponds to a MSN.}

\subsection{\textcolor{black}{Experimental realization with non-linear electronic circuits}}

\begin{figure}
\begin{center}
\includegraphics[scale=0.10]{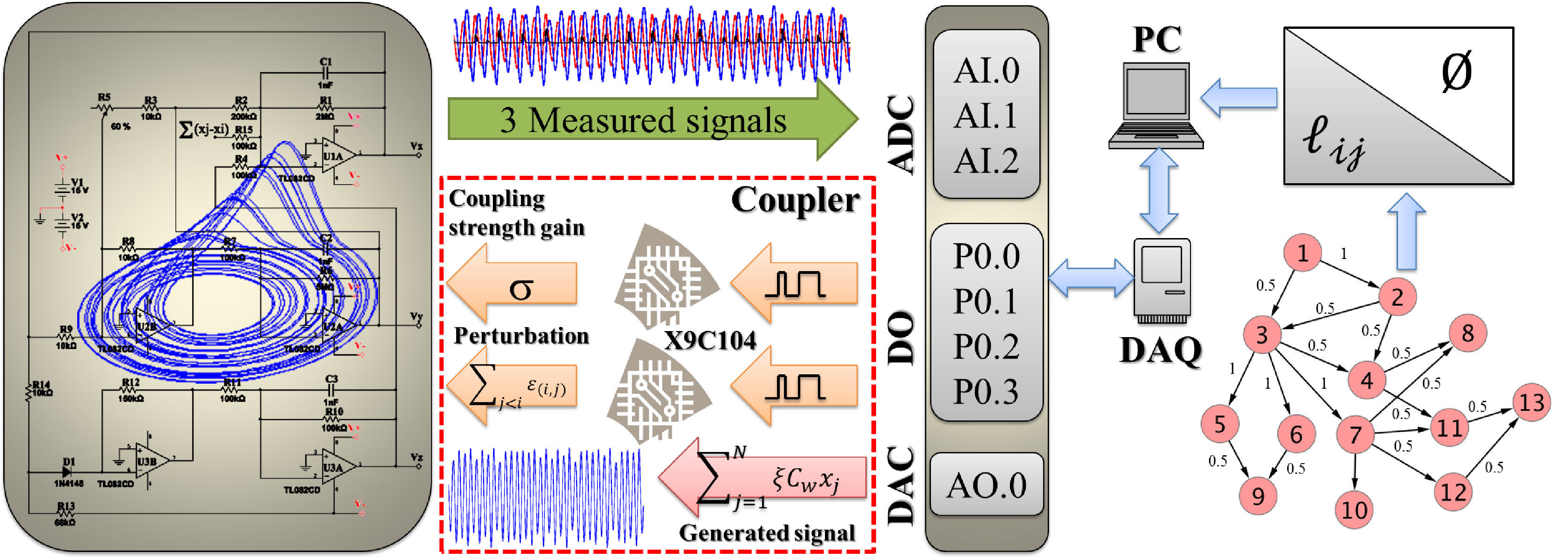}
\end{center}
\caption{ {\sf \bf Experimental setup.}
Electronic implementation of a network of R\"{o}ssler oscillators. The three variables of a R\"ossler chaotic circuit are recoded by means of an analog-to-digital card (ADC)
and recorded in a computer. The same circuit is used to simulate all nodes of the network, whose coupling matrix (and weights) is sent from the computer to the circuit through a digital-to-analog card (DAC). A digital line (DO), controls the coupling strength and the gain product of perturbations by means of digital potentiometers. See the Appendix for a detailed description of the circuit parameters.
}
\label{fig2}
\end{figure}

The procedure described above can be applied experimentally to arbitrary large networks of electronic circuits by adapting the methodology first developed in Ref. \cite{pisarchik2009}. In simple terms, a large unidirectional network of nonlinear circuits is obtained by the sequential recording of the time series of successive layers of neighbors and the weighted reinjection of the data from previous layers using just one electronic circuit. The technical details are discussed in the Appendix, and an illustration of the experimental setup is provided in Fig. \ref{fig2}.

Our experimental system is a network of piecewise linear R\"{o}ssler-like electronic circuits, as illustrated in Fig. \ref{fig2}. The dynamics of node $i$ is given by the following equations \cite{CP}:
\begin{equation}
\begin{array}{l}
\dot{x}=-\alpha_{1_{i}} \left(x_i+\beta y_i+\Gamma z_i+\sigma \psi \sum^{N}_{j=1} \mathcal{L}_{ij} x_{j}\right), \\
\dot{y}=-\alpha_{2_{i}} \left(-\gamma x_i+[1-\delta] y_i\right), \\
\dot{z}=-\alpha_{3_{i}} \left(-g(x_i)+z_i\right),
\label{Rossler1}
\end{array}
\end{equation}
\noindent where $x$, $y$ and $z$ are the oscillator state variables. The piecewise linear function $g(x)$ defined as
\begin{equation}
g(x_i)=\left\{
\begin{array}{cc}
0 & x_i\leq 3, \\
\mu \left( x_i-3\right)  & x_i>3,%
\end{array}%
\right.  \label{geq}
\end{equation}%
introduces the nonlinearity in the system that leads to a chaotic behavior. The parameter values are $\alpha_1 = 500$, $\alpha_2 = 200$, $\alpha_3 = 10000$, $\beta=10$, $\Gamma=20$, $\gamma=50$, $\delta=10.0402$, $\psi = 20$ and $\mu=15$. The parameter $\sigma$ is the coupling strength, which can be adjusted. For a detailed study of this dynamical system, whose attractor is sketched in Fig. \ref{fig2}, see Refs. \cite{CP, Pis, PRL}. Concerning the initial topologies of the networks, we \textcolor{black}{next} report results obtained with scale-free networks, \textcolor{black}{as this is a very relevant case from an experimental point of view. The network size considered is $N=200$}. 

\subsection{\textcolor{black}{Experimental results and a comparison with MSF predictions}}

\begin{figure}
{ \begin{center}
\includegraphics[scale=0.40]{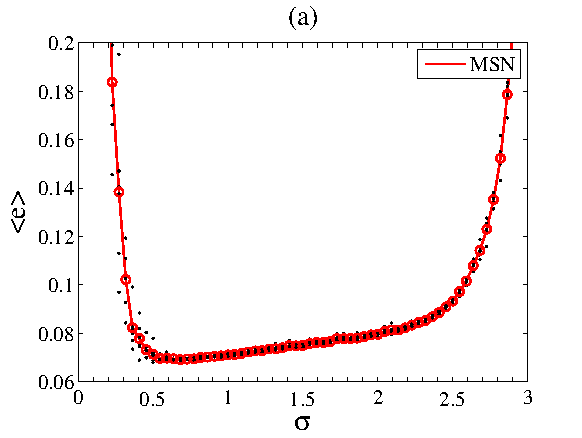}
\includegraphics[scale=0.40]{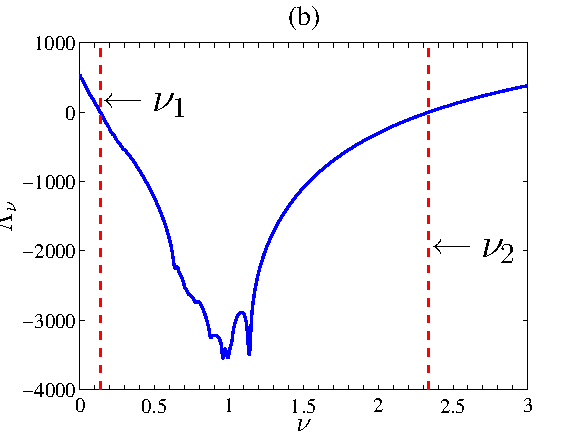}
\end{center}
}
\caption{ {\sf \bf Synchronization error and MSF of the system.}
(a) Synchronization error  $\langle e \rangle$ of experimental MSN for 6 different realizations (black dots) and the average across realizations (red continuous line). (b) Numerically obtained MSF for the system described in Eq. \ref{Rossler1}. Dashed lines indicate the values of $\nu_1=0.137$ and $\nu_2=2.337$.
}
\label{fig3}
\end{figure}

We capture the dynamics of the $x(t)$ variable of each circuit and compute the overall synchronization error in the
network, $\langle e \rangle=\sum_{i,j}{D_{x_i x_j}}/N^2$, where $D_{x_i x_j}=\langle | x_i(t)-x_j(t)| \rangle$ and the angular brackets stand for a time averaging. Times series have a length of $50000$ points after a transient of $5000$ is disregarded. The synchronization error $\langle e \rangle$ is shown in Fig. \ref{fig3} (a), where points corresponding to six different dynamical realizations of the same MSN are shown, and the continuous line is the average. If we focus on the lower values of $e$ to study when the system is the closest to complete synchronization, we can see that the system is synchronized from, roughly, $\sigma = 0.40$ to $\sigma = 2.40$. The fact that the system becomes unsynchronizable for low and high values of $\sigma$ indicates that the R\"{o}ssler system coupled through the $x$ variable is a class III system. This fact is confirmed when the MSF corresponding to Eq. \ref{Rossler1} is calculated numerically, as shown in Fig. \ref{fig3} (b). \textcolor{black}{This calculation is performed by linearizing the equations of motion according to the reasoning described in \cite{pecora1998}. The calculation of the maximum Lyapunov exponent, which is based on the time evolution of a vector in the tangent space that is periodically renormalized, follows the method proposed in \cite{benettin1980}. The justification for the use of such a method to calculate the Lyapunov exponent of a piecewise-linear system representing the macroscopic behavior of electronic systems, such as the one given by Eq. \ref{Rossler1}, is provided in \cite{sevilla2015}.} 

\begin{figure}
{ \begin{center}
\includegraphics[scale=0.40]{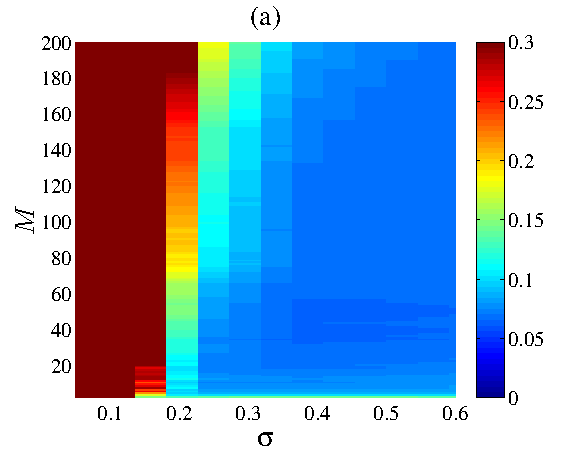}
\includegraphics[scale=0.40]{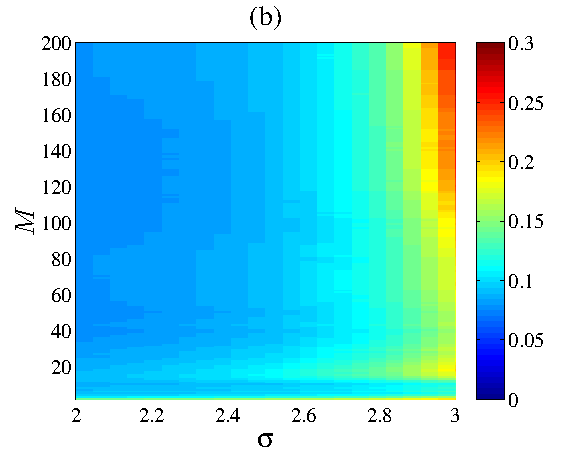}
\end{center}
}
\caption{ {\sf \bf  Subgraph synchronization error $\langle e \rangle_{sub}$ around the values of $\nu_1$ and $\nu_2$.}
(a) Subgraph synchronization error $\langle e \rangle_{sub}$ around the first zero $\nu_1=0.137$ of the MSF for different subgraph sizes $M$. \textcolor{black}{(b)} The same as in (a) but around the second zero $\nu_2=2.337.$
}
\label{fig4a}
\end{figure}

Since the MSN has $\lambda_2 = \cdots = \lambda_N = 1$, one can identify $\nu$ with $\sigma$. We can see that while the second zero $\nu_2=2.337$ is close to the upper boundary of the synchronization region in the experimental results of Fig. \ref{fig3} (a), the first zero $\nu_1=0.137$ is significantly smaller than the lower boundary in relative terms. To further investigate this discrepancy we have obtained the {\it subgraph synchronization error} $\langle e \rangle_{sub}$ around the values of $\nu_1$ and $\nu_2$, computed as the synchronization error within the subgraph given by a node and the next $M$ nodes in increasing order as given by the node labelling. Fig. \ref{fig4a} (a) and (b) show the values of $\langle e \rangle_{sub}$ as a function of the subgraph size $M$ and the coupling strengths surrounding the corresponding values of $\nu_1$ and $\nu_2$. We can observe how at the boundaries of the synchronization region, $\langle e \rangle_{sub}$ increases with $M$, which indicates that the experimental error is propagating through the network.

We conclude that the system size dependence of the synchronization region that is found experimentally can be attributed to the accumulation of synchronization error across the network layers.
While \textcolor{black}{one could be tempted to attribute this effects to} the 12-bit resolution of the analog-to-digital acquisition card (ADC) combined with a sampling of 100 $kS/s$ (kilosamples per second)\textcolor{black}{, which} leads to truncation errors which are amplified due to the chaotic dynamics of the system, \textcolor{black}{or to drifts in the electronic component properties that make the nodes effectively more different as time goes by, indeed these experimental uncertainties are not at the origin of these effects. The fact that the synchronization error propagates in a very similar way in numerical data (not shown here) obtained from the numerical integration of the system with a Runge-Kutta algorithm (where resolution errors are much smaller), indicates that this propagation exists also in the absence of those experimental issues. Moreover, the pattern of the synchronization error propagation is clear enough (both in the numerical and the experimental results) to suggest a quite compelling dynamical origin, as we now explain.}

\begin{figure}
{ \begin{center}
\includegraphics[scale=0.28]{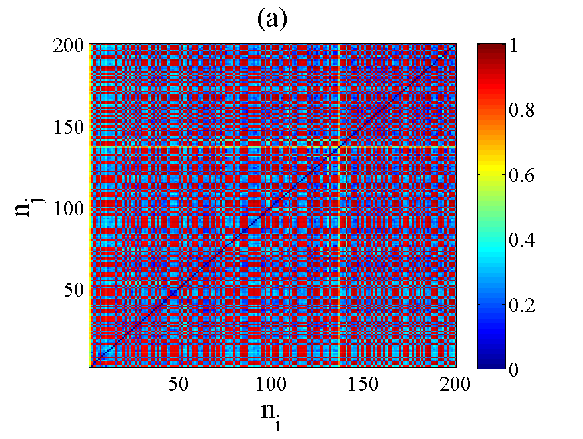}
\includegraphics[scale=0.28]{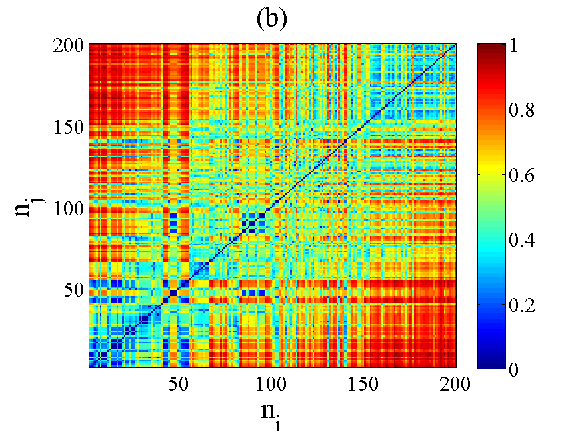}\\
\includegraphics[scale=0.28]{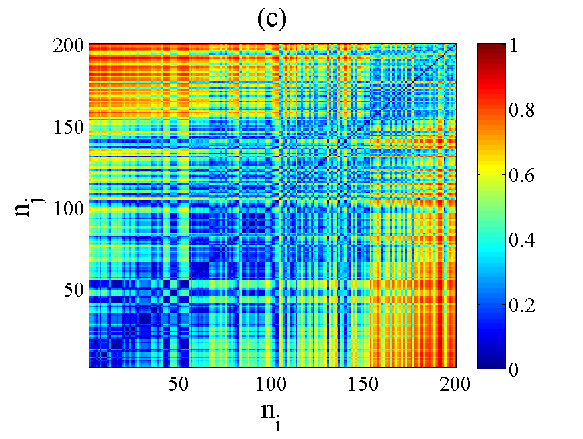}
\includegraphics[scale=0.28]{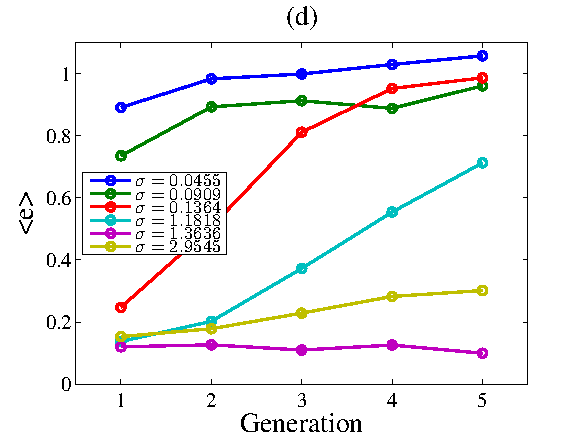}\\
 \end{center}
}
\caption{ {\sf \bf {Pairwise synchronization error for different coupling strengths}.}
Pairwise synchronization matrices for  \textcolor{black}{(a) $\sigma = 0.0455$, (b) $\sigma = 0.1364$, (c) $\sigma = 1.1818$. The values of the coupling strength $\sigma$ in (c) fall within the synchronization region.} In these cases, the colored maps show the formation of clusters, as indicated by the rise of the pairwise error with increasing distance between the nodes. When the dynamics is not synchronous, (a), the effect of the error accumulation is negligible. In all cases the matrices have been normalized by its largest element. \textcolor{black}{In panel (d) we show the average pairwise synchronization error between node $1$ and its first, second, third... neighbors for the several $\sigma$ values (including those shown in the other panels). The propagation is monotonically increasing in and close to the synchronization region.} }
\label{fig4}
\end{figure}

\textcolor{black}{Figure \ref{fig4} (a), (b) and (c) illustrates the propagation of the error by plotting the pairwise error between all pairs of nodes for different values of $\sigma$: from values lower than $\nu_1$ to higher values, moving across the synchronization region.} The synchronization error of these maps is normalized by the maximum achieved at any element of the matrix to facilitate comparisons between them. While in the completely asynchronous regime the propagation of the error  does not have any significant effect, it is clear that within and around the synchronized regime the effect is quite visible, since nodes that are topologically further apart have a significantly larger synchronization error between them. \textcolor{black}{In panel (d), the average synchronization error between node $1$ and the nodes that are at a given topological distance from it (its first neighbors being separated a distance $1$, its second neighbors a distance $2$,...) is reported, which further confirms this finding, as do numerical results, where a similar increase in the error as a function of the topological distance is observed in almost synchronous states (while perfectly synchronous states, which cannot exist in the experiment, simply show zero error for all topological distances). All this points to a dynamical origin of the propagation of errors. Close to the synchronization region, or even within it in an experimental realization (where perfect synchronization is never achieved), as node 1 tries to impose its state upon its first neighbors, it fails to do so, leaving its first neighbors in a nearby dynamical state. The same thing happens between these nodes and those neighbors to which they are linked, and therefore the discrepancies with respect to the state of node $1$ (or any other reference node) are only expected to increase as one moves across the network layers to more distant nodes. In short, our results suggest that near the perfectly synchronized state, topologically close nodes are more synchronized than distant nodes by the unidirectional coupling structure of the MSN tree.}
%Why this kind of error accumulation affects more seriously the position of the effective $\nu_1$ than that of $\nu_2$ (compare Fig. \ref{fig3} (c) and (d)) is something for which at present we do not have a convincing argument.

\section{Topologically perturbed maximally synchronizable network}

So far we have been dealing with ideal MSN topologies, where $R = 1$ holds exactly (within the small experimental errors because the electrical
components have a 5\% tolerance, the temperature is not a constant, and the recording equipment has a
finite precision due to the analog-digital conversion). Nevertheless, in experimental realizations, or real technological networks, noise is always at play and one cannot expect such idealized model to reflect realistic conditions in less carefully controlled environments. Not only is the intrinsic {\it dynamical} noise of the system present, but also deviations from the optimal topology can be expected (i.e., link weights could be affected by perturbations). In this section we propose some theoretical estimates about effect of {\it topological noise} on synchronizability \textcolor{black}{in the first subsection}, and \textcolor{black}{then, in the second subsection,} validate these predictions experimentally. Moreover, as frequently a very detailed knowledge of the topology may not be readily available, or may be extremely cumbersome to deal with if the network is very large, we favor an approach to the study of this issue based on a very limited knowledge of the system topology.

\subsection{\textcolor{black}{Estimating the effect of topological noise on network synchronizability}}

Let us assume the link weights in the MSN are uniformly perturbed with additive noise. The perturbation that affects the link between nodes $i$ and $j$ (assuming $i$ and $j$ are connected) is denoted $\epsilon_{(i,j)}$. The Laplacian matrix of a generic perturbed MSN is therefore
\begin{equation}
\mathcal{L'}= \left (
 \begin{array}{l l l l }
  0 & 0 & \cdots & 0 \\
  -1 - \epsilon_{(2,1)} & 1 + \epsilon_2 & \cdots & 0 \\
  -1/k^{in}_3 - \epsilon_{(3,1)} (0) & -1/k^{in}_3 - \epsilon_{(3,2)} (0) & \cdots & 0 \\
  \vdots  & \vdots & \ddots & \vdots  \\
  -1/k^{in}_N - \epsilon_{(N,1)}(0) & -1/k^{in}_N - \epsilon_{(N,2)} (0) & \cdots & 1 + \epsilon_N 
 \end{array} \right )
\end{equation}
where $\epsilon_i \equiv \sum_{j<i} \epsilon_{(i,j)}$ considering  $\epsilon_{(i,j)} = 0$ for unconnected $i$ and $j$, and $\epsilon_2 = \epsilon_{2,1}$. The Laplacian spectrum is given by $\textrm{diag}(\mathcal{L'}) = \{0,1+\epsilon_2,1+\epsilon_3,\ldots,1+\epsilon_N\}$. For a given perturbed topology (i.e., for a given realization of $\epsilon_{(i,j)}$ for all $i$ and $j$), we define $\epsilon_\textrm{max} = \textrm{max}\{\epsilon_2,\epsilon_3,\cdots,\epsilon_N\}$, and $\epsilon_\textrm{min}$ analogously. The perturbed graph eigenvalues that affect the synchronizability properties of the system are therefore $\lambda_2^\textrm{pert} = 1 + \epsilon_\textrm{min}$ and $\lambda_N^\textrm{pert} = 1 + \epsilon_\textrm{max}$.

In order to be able to make some concrete predictions, one has to assume a particular distribution for the noise terms. Let us make the reasonable assumption that $\epsilon_{(i,j)}$ for all $i$ and $j$ are independent and identically distributed Gaussian random variables of zero mean and standard deviation $\Sigma$, which we denote $\mathcal{G}(0,\Sigma)$ (our predictions can be, in principle, adapted to other probability densities). Here, $\Sigma$ plays the role of the topological noise strength. From the well-known properties of the sums of Gaussian random variables, $\epsilon_i$ for $i=0,\cdots,N-1$ are therefore random variables distributed according to $\mathcal{G}(0,\sqrt{k_i} \Sigma)$. We assume the only knowledge on the topology we have access to is a suitably defined {\it typical degree} $k_\textrm{typ}$ of the network. In regular random graphs such as Erd\"{o}s-R\'enyi graphs, it makes sense to consider the mean degree, $k_\textrm{typ} \sim \langle k \rangle$; however, in scale-free networks where $\langle k \rangle$ may be so much affected by the very large connectivity of some of the hubs, the median of the degree distribution may be a better choice $k_\textrm{typ} \sim \tilde{k}$. This very rudimentary knowledge can prove quite useful in giving estimates to the effect of noise, if one has also some information about the noise strength. A typical node is perturbed by a noise $\bar{\epsilon}$ distributed according to $\mathcal{G}(0,\sqrt{k_\textrm{typ}} \Sigma)$, and we define the probability to obtain a value of $\bar{\epsilon}$ that is larger than $\Delta>0$ as
$p_{\Delta} \equiv (1/\sqrt{2\pi k_\textrm{typ}}\Sigma) \int_{\Delta}^\infty dx\, \exp{\left(-x^2/2 k_\textrm{typ} \Sigma^2\right)}.$
Obviously, the probability that $\bar{\epsilon}$ is smaller than $-\Delta$ is also $p_{\Delta}$. As the noise affecting different nodes is stochastically independent, if $p_{\Delta} \geq 1/(N-1)$ holds, we may expect to have on average at least one node with noise intensity equal or greater than $\Delta$ in absolute value.

The procedure we propose consists in inverting the previous chain of reasoning. For a network of size $N$ and typical degree $k_\textrm{typ}$, one first obtains the value $\Delta>0$ such that the inequality above is exactly satisfied as an equality, $p_\delta = 1/(N-1)$. We denote this value as $\delta$, while $p_\delta$ denotes the probability that a random variable distributed according to $\mathcal{G}(0,\sqrt{k_\textrm{typ}} \epsilon)$ takes on values larger than $\delta$. We expect that there will be on average one node in the network that is affected by a noise term greater than $\delta$ and also another one that is affected by a noise term smaller than $-\delta$. Thus, we expect $\epsilon_\textrm{max} \simeq \delta$ and $\epsilon_\textrm{min} \simeq - \delta$, and therefore $\lambda_2^\textrm{pert} \simeq 1 - \delta$ and $\lambda_N^\textrm{pert} \simeq 1 + \delta$. Given the simplicity of the approximation, which is based only on knowledge of $N$, $k_\textrm{typ}$ and $\epsilon$, one cannot expect the estimates that result from it to be very precise. Nevertheless, we will see \textcolor{black}{in the next subsection} that it usefully predicts the effect of topological noise on the network synchronizability quite satisfactorily.

\subsection{\textcolor{black}{Experimental observation of the effect of topological noise on network synchronizability}}

One has to consider that when a relatively weak noisy signal is injected to an electronic system in a controlled fashion to study its effects, the precise noise strength existing in the system is hard to determine, due to the intrinsic (dynamical) noise present in electronic circuitry and other experimental equipment that somehow add up to the noisy signal. A similar remark can be made about topological noise. The experimental study of the effects of topological noise on the synchronizability was designed with $N=50$ piecewise R\"ossler oscillators coupled through a MSN scheme obtained from a scale-free network following the procedure described above (see Fig \ref{fig1}). We average $20$ independent realizations for coupling strengths around the area of the first zero $\nu_1$ of the MSF. Two Gaussian noise strengths are considered, $\Sigma = 0.015$ and $0.030$, together with the case without topological noise $(\Sigma = 0.000)$.

\begin{figure}
{ \begin{center}
\includegraphics[scale=0.50]{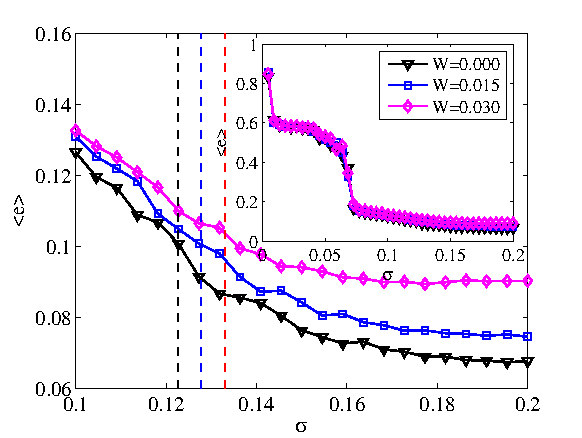}
\end{center}
}
\caption{ {\sf \bf {Synchronization error around the first MSF zero for different topological noise strengths}.}
Experimental study of the synchronization error for $\Sigma = 0.000, 0.015, 0.030$ with $N=50$. The results here reported are averages of 20 independent realizations. \textcolor{black}{Inset: same results for the full coupling range that has been experimentally explored ($\sigma \in [0, 0.2]$}).
}
\label{fig5}
\end{figure}

As we can observe in Fig. \ref{fig5}, the onset of synchronization is seen to occur for larger $\sigma$ as the noise strength is increased, as one would expect by the fact that the smallest eigenvalue is expected to decrease with respect to the MSN case (see the explanation above). Moreover estimates of the effect of noise on the onset of synchronization based on the previous reasoning were also obtained (using the median of the degree distribution $\tilde{k}$ as the only topological information, which plays the role of $k_\textrm{typ}$), showing that indeed the method outlined above is useful in predicting the effects of topological perturbations on the network synchronizability. In the experiment, it is difficult to establish a well-defined onset of synchronization, as complete synchronization is never perfectly reached, and the level of synchronization achieved for any $\sigma$ is dependent on $\Sigma$. We consider the synchronization threshold to be $\langle e \rangle = 0.1$, which indicated with a black vertical dashed line for $\Sigma = 0$.  The associated increase in the coupling strength $\sigma$ that is needed to reach the first zero of the MSF (considering $\sigma = \nu_1/(1-\delta)$) is $4\%$ (for $\Sigma = 0.015$) and $8.4\%$ (for $\Sigma = 0.030$). These estimates are indicated with vertical lines in Fig. \ref{fig5}

%%%%%%%%%%%%%%%%%%%%%%%%%%%%%%%%%%%%%%%%%%%%%%%%%%%%%%%%%%%%%%%%%%%
\section{Conclusions}
%%%%%%%%%%%%%%%%%%%%%%%%%%%%%%%%%%%%%%%%%%%%%%%%%%%%%%%%%%%%%%%%%%%

In summary, we have presented a strategy applicable to a generic connected and undirected network, which transform it into
its directed, {\it maximally synchronizable}, configuration. Our approach, indeed, allows tuning the directions and weights of the links of any given network, so as to reach the lowest possible eigenratio $\lambda_N/\lambda_2$, which guarantees on its turn maintenance of the stability of the synchronous state across the longest possible range of coupling strengths
in class III systems.

Far from constituting a merely theoretical proposition, we have proved
the feasibility, effectiveness and robustness of the method by means of an experiment with chaotic electronic oscillators, 
this way validating the technique also
in the case of non perfectly identical systems.
Furthermore, our experiment allowed to monitor the propagation of the synchronization error throughout the network, as the system approaches the synchronization boundaries, as well as to analyze the effects of deviations from the optimal topology (the maximally synchronizable configuration) elucidating the 
crucial interplay between {\it topological noise} and the coupling strength of the whole network.

Our results are therefore useful as a guide for implementation of the method in other networked dynamical systems, as well as for engineering topologies of specific subgraphs of a larger network wherein one would desire synchronization dynamics to be supported and kept across a large variability of global parameters, such as the coupling strength.

%%%%%%%%%%%%%%%%%%%%%%%%%%%%%%%%%%%%%%%%%%%%%%%%%%%%%%%%%%%%%%%%%%%
\section{Acknowledgements}
%%%%%%%%%%%%%%%%%%%%%%%%%%%%%%%%%%%%%%%%%%%%%%%%%%%%%%%%%%%%%%%%%%%
The authors acknowledge J. L. Echenaus\'ia-Monroy, V. P. Vera-\'Avila, J. Moreno de Le\'on, C. Hapo and P.L. del Barrio for assistance in the laboratory, and the support of MINECO (FIS2012-38949-C03-01 and FIS2013-41057-P). \textcolor{black}{One anonymous referee is acknowledged for having provided valuable advice that has influenced our understanding of the origin of the propagation of the synchronization error, and helped us improve the manuscript in several ways.} The authors also acknowledge the computational resources, facilities and assistance provided by the Centro computazionale di  RicErca sui Sistemi COmplessi (CRESCO) of the Italian National Agency for New Technologies, Energy and Sustainable Economic Development (ENEA). R.S.E. acknowledges Universidad de Guadalajara, CULagos (Mexico) for financial support (PRO-SNI-2015/228069, PROINPEP/005/2014, UDG-CONACyT/I010/163/2014) and CONACyT (Becas Mixtas MZO2015/290842). D.-U. Hwang acknowledges  National Institute for Mathematical Sciences (NIMS) funded by the Ministry of Science, ICT \& Future Planning (A21501-3).

%%%%%%%%%%%%%%%%%%%%%%%%%%%%%%%%%%%%%%%%%%%%%%%%%%%%%%%%%%%%%%%%%%%
\section*{Appendix - Electronic implementation of an acyclic complex network}
%%%%%%%%%%%%%%%%%%%%%%%%%%%%%%%%%%%%%%%%%%%%%%%%%%%%%%%%%%%%%%%%%%%

The experimental design is similar to that reported in \cite{pisarchik2009}. It consists of an electronic array, a card of acquisition/generation data, and a computer, as shown in Fig. \ref{fig2}. The whole experimental process is controlled from a virtual interface developed in Labview 2012, which can be considered as a state machine adapted to the study of complex networks with unidirectional fixed topology. As a first step, it is necessary to load the desired network topology so that the interface can define the links connecting the nodes. Then, the time series of all variables of an isolated R\"ossler oscillator circuit ($\sigma = 0$) is recorded via an analogue-digital converter (ADC), with a  sampling rate of 100 $kS/s$ (kilosamples per second) and a 12 -bit resolution. This circuit is called node 1. Once the time series of the three variables of the system ($x$,$y$ and $z$, as shown in Eq. \ref{Rossler1}) have been stored, they are converted into electrical signals through a digital-analogue converter (DAC) and subsequently reinjected to the electronic circuit, which is called node 2, via a coupler circuit (XDCP) and the response of the circuit is stored. Note that node 1 and node 2 are, in fact, the same electronic circuit, but the inputs and outputs of both systems are different. Thus, by using just one oscillator, the coupling is effectively made between a pair of nodes through the ADC and we obtain the dynamics of nodes $1$ and $2$ in the network. For the rest of the neighbors of node $1$ (if they exist), the process is repeated as for node $2$. Next, the same procedure is applied for the second neighbors, third neighbors, etc. The variables of all nodes of the network are stored in the computer since, in principle, they could be use as input signals of subsequent nodes. The only requirements are, first, that the network must be acyclic and, second, that the output of a node must be injected into a node with higher numbering, both conditions fulfilled by the MSN. Importantly, the interface digitally performs the sum of the different signals entering a given node, after multiplying each of these by the appropriate link weight ($W_{ij}$) and network strength ($\sigma$) according to the connectivity matrix previously defined. %\textcolor{black}{ In this part of the process it is assumed that $R = 1$ in case the MSN pertubed is only necessary to add a second digital potentiometer, so that depending on the own eigenratio of adjusting the circuit gain in the coupling step node. This is achieved medianta circuit including a cascade consisting of two inverting amplifiers where the first of these has a variable gain (X9C104, 10\% tolerance) so that can be increased or decreased, the second has unit gain and works only for adjusting the sign of signal. Note that this this interface is fully autonomous and therefore the gain of weight $1/k^{in}_i$ will be discretized 100 possible values, so it is necessary to define a minimum and maximum disturbance, so that by varying the gain feedback amplifier we can change this weight for each node}.
It is worth mentioning that the coupling circuit has a digital control stage, which allows changing the coupling between nodes autonomously, thus conducting experiments with networks with a large number of elements is possible and only limited by the storage capacity on the hard drive of the computer.

\end{document}